\documentclass{article}
\usepackage{spconf,amsmath,graphicx,hyperref}
\usepackage{subcaption} 

\def\bilstmof{{\sc BiLSTM-of }}
\def\onedcnnof{{\sc 1D-CNN }}
\def\vitof{{\sc ViT-Ave }}

\newtheorem{my_note}{Remark}

\def\all{{C_{+}}}
\def\model#1{{\sc model(#1)}}

\title{SAND Challenge:  Four Approaches for Dysarthria Severity Classification }
\name{Gauri Deshpande, Harish Battula, Ashish Panda, Sunil Kumar Kopparapu}
\address{TCS Research, Tata Consultancy Services Limited, India}

\begin{document}
\maketitle

% Abstract
\begin{abstract}
This paper presents a unified study of four distinct modeling approaches for classifying dysarthria severity in the Speech Analysis for Neurodegenerative Diseases (SAND) challenge. All models tackle the same five-class classification task (ALS patients rated 1–4 in severity, and 5 for healthy controls) using a common dataset of ALS patient speech recordings. We investigate: (1) a \vitof{} method leveraging a Vision Transformer on spectrogram images with an averaged-loss training strategy, (2) a \onedcnnof{} approach using eight 1-D convolutional neural networks (CNNs) with majority-vote fusion, (3) a \bilstmof{} approach using nine BiLSTM models with majority-vote fusion, and (4) a Hierarchical XGBoost ensemble that combines glottal and formant features through a two-stage learning framework. Each method is described, and their performances on a validation set of 53 speakers are compared. Results show that while the feature-engineered XGBoost ensemble achieves the highest macro-F1 (0.86), the deep learning models (ViT, CNN, BiLSTM) attain competitive F1-scores (0.64–0.70) and offer complementary insights into the problem. We discuss the trade-offs of each approach and highlight how they can be seen as complementary strategies addressing different facets of dysarthria classification. In conclusion, combining domain-specific features with data-driven methods appears promising for robust dysarthria severity prediction. 

\end{abstract}


% Introduction
\section{Introduction}

Dysarthria is a motor speech disorder common in neuro-degenerative diseases such as Amyotrophic Lateral Sclerosis (ALS). The Speech Analysis for Neuro-degenerative Diseases (SAND) challenge \cite{SAND2026TaskOne} at ICASSP 2026 focuses on automatic classification of dysarthria severity into five levels. Task \#1 of this challenge asks participants to predict the severity class (“ALSFRS-R” score category) for each speaker’s voice, given a fixed set of short utterances (spoken vowels and syllables). Class labels range from 1 (most severe dysarthric speech) to 4 (milder dysarthria in ALS patients), and 5 for healthy control speakers. 

All approaches in this study use the same dataset provided in SAND challenge. The dataset contains recorded utterances from 219 ALS patients for training and 53 speakers for validation. Each speaker provides 8 specific utterances: five sustained phonations (vowels A, E, I, O, U) and three repetitive rhythmic syllables (KA, PA, TA). These utterances capture different aspects of speech production (vowel phonation versus articulatory rhythm). The classification task is challenging due to severe class imbalance – for example, only 4 speakers are labeled Class 1 (most severe) while 86 are Class 5 (healthy) in the training set. This imbalance necessitates strategies like data augmentation and weighted loss to avoid biasing toward the majority class. Additionally, there is a gender imbalance, with male-to-female ratios of $1.28$ in training and $1.30$ in validation sets (see Table \ref{tab:data_distribution}).

In this paper, we consolidate four complementary models developed for SAND Task \#1, integrating our findings into a single cohesive report. Despite differing methodologies, all four aim to maximize classification accuracy on the same task and dataset. By unifying their perspectives, we provide a comprehensive view of how diverse techniques, ranging from deep learning on raw spectrograms to machine learning on engineered features, can contribute to the dysarthria severity classification problem. The following sections describe each approach’s methodology, followed by a comparison of their performance and a discussion on their complementary strengths. 

% The SAND Challenge  aims to classify dysarthria severity in ALS patients and healthy controls using speech signals. Task \#1 involves five-class classification at time zero (first assessment). The dataset comprises eight utterances per speaker (phonation: A, E, I, O, U; rhythm: KA, TA, PA) from $219$ training and $53$ validation speakers. A significant challenge in this task is the severe class imbalance, as shown in Table \ref{tab:data_distribution}. For example, Class $1$ has only $4$ speakers in the training set compared to $86$ speakers in Class $5$. 

\begin{table}[!htbp]
\centering
\caption{Gender distribution across classes, SAND Task \#1 dataset.}
\label{tab:data_distribution}
\begin{tabular}{|c|ccc|c|ccc|}
\hline
\multicolumn{4}{|c|}{\textbf{Training Baseline}} & \multicolumn{4}{c|}{\textbf{Validation Baseline}}\\
\hline
\textbf{Class} & F & M & Total & \textbf{Class} & F & M & Total \\
\hline
$1$ & $3$ & $1$ & $4$ & $1$ & $1$ & $1$ & $2$\\
$2$ & $12$ & $10$ & $22$ & $2$ & $3$ & $1$ & $4$\\
$3$ & $16$ & $29$ & $45$ & $3$ & $5$ & $7$ & $12$\\
$4$ & $24$ & $38$ & $62$ & $4$ & $4$ & $10$ & $14$\\
$5$ & $41$ & $45$ & $86$ & $5$ & $10$ & $11$ & $21$\\
\hline
\textbf{Total} & $96$ & $123$ & $219$ & \textbf{Total} & $23$ & $30$ & $53$ \\
\hline
\multicolumn{4}{|c|}{Male/Female ratio = $1.28$} & \multicolumn{4}{c|}{Male/Female ratio = $1.30$} \\
\hline
\end{tabular}
\end{table}

% Methodology
\section{Methodology}
We developed four different models to address the five-class dysarthria classification. Each approach leverages a unique modeling technique and fusion strategy for the multiple utterances per speaker. In the following, we detail each approach: \vitof{}, Hierarchical XGBoost, \onedcnnof{}, and \bilstmof{}. 

\subsection{ViT Model with Averaged Loss (\vitof)}

\begin{figure}[!htbp]
    \centering
    \includegraphics[width=0.30\textwidth]{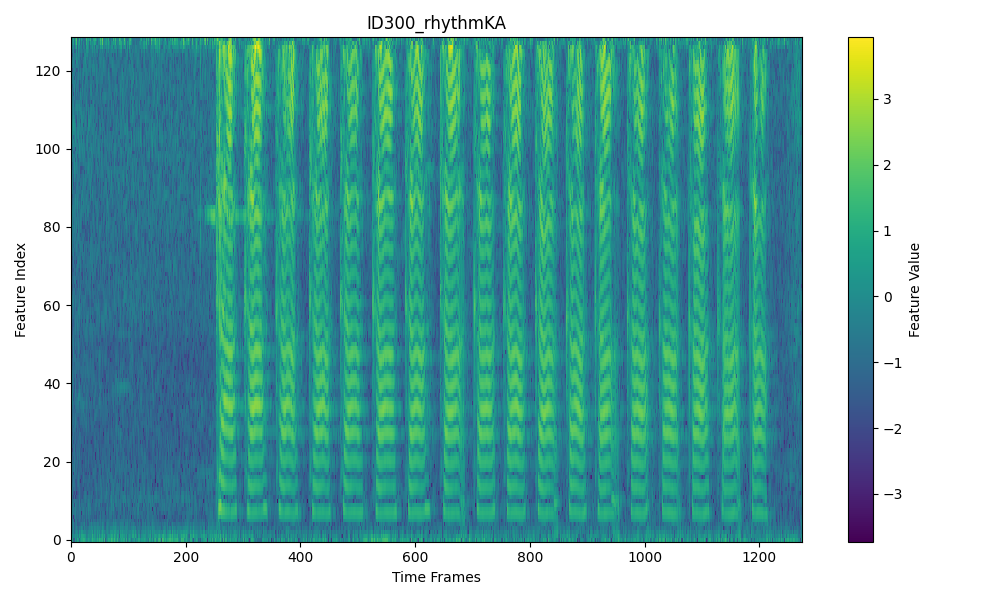}
    \caption{Spectrogram of ID300 for the rhythm KA.}
    \label{fig:spectrogram-300ka}
\end{figure}

The \vitof{} approach uses a vision transformer model to classify dysarthria severity from spectrogram images. We started with a pre-trained Vision Transformer (ViT-B16) model (originally developed for image recognition) and fine-tuned it on the speech spectrogram data. Each audio utterance (vowel or syllable) was converted to a 2D spectrogram image (see Fig. \ref{fig:spectrogram-300ka}) using {\tt librosa}; images were standardized to $512 \times 256$ pixels representing the time-frequency content. Basic data augmentation was applied to these spectrogram images during training (random horizontal flips, rotations, and color jitter for brightness/contrast/saturation) to increase variability. 

A key novelty of \vitof{} is how it handles multiple utterances per speaker. During training, we computed the loss for each of a speaker’s 8 (5 phonations, 3 rhythm) spectrograms and then averaged the loss across all utterances of that speaker before back-propagation. This averaged loss approach aligns with the assumption that a speaker’s overall severity label should be reflected consistently across all their utterances. It effectively treats the 8 utterances as a set, stabilizing training by not over-weighing any single utterance. In the inference (validation) phase, we feed all 8 spectrograms of a speaker through the ViT model and then average the predicted class probabilities of all utterances to make the final decision. This probability averaging (late fusion) is analogous to an ensemble vote, but weighted by confidence, and was found to improve robustness. 

\subsubsection{Architecture} The \vitof{} model is a  PyTorch pretrained model ViT-B16 \cite{50650} transformer (12 layer, 768 dim hidden) with its final classification head replaced by a new fully connected layer tuned for 5 classes. By leveraging the ViT’s strength in image analysis, the model can capture fine-grained spectral patterns in the voice recordings. 

\subsubsection{Result} On the validation set (53 speakers), the \vitof{} achieved a macro-averaged F1 score of 0.68 (68.09\%). Notably, this approach demonstrated that even with relatively few training samples, a pre-trained transformer can be re-purposed for audio classification when combined with appropriate augmentation and a fusion strategy. We observed that data scarcity and imbalance caused a noisy validation curve (high variance in F1 across epochs), indicating the model was somewhat sensitive to training fluctuations. Nonetheless, \vitof{} provided a solid baseline, outperforming a trivial majority-class predictor by leveraging spectral image-based features and multi-utterance averaging. 

%  We replace the final layer with a fully connected layer which is finetuned for $5$ class classification. The spectrograms of each speech utterance is generated by librosa and is saved as a $512x256$ image. During training, for each speaker ID all the $8$ spectrogram images are loaded and augmentation techniques are applied. We have used random horizontal flip, random rotation and colour jitter (brightness=$0.2$, contrast=$0.2$, saturation=$0.2$) followed by normalization. Each of the spectrogram is input to the model and loss is computed and it is averaged over all the eight images for each speaker ID. This averaged loss mimicks the process of grading each speakers class through averaging over all speech utterances.

% During prediction phase, for validation set and test test, we again load all the eight spectrograms for a given speaker ID. Using softmax function probality of for each class is generated for each of the spectrograms. These probabilities are summed across all the eight spectrograms to find the most likely class. 

\begin{table*}[!htbp]
\centering
\caption{Configuration of eight XGBoost models for hierarchical approach.}
\label{tab:model_config}
\begin{tabular}{|c|c|c|c|c|c|c|}
\hline
\textbf{Model \#} & \textbf{Group (M-F)} & \textbf{Classification} & \textbf{SC} & \textbf{$N_{est}$} & \textbf{$F_l$} (ms) & \textbf{$Audio_{d}$}\\ \hline
$1$ &  F &   $3$ vs  $4$ \& $5$ & A & $200$ & $100$  & Full\\ \hline
$2$ &  F &   $4$ vs   $5$ & KA & $100$ & $100$   & Initial $20$  s\\ \hline
$3$ & $< 60$- M &   $3$ vs  $4$ \& $5$ & U & $100$ & $50$   & Later $10$  s\\ \hline
$4$ & $< 60$- M &   $4$ vs   $5$ & O & $100$ & $500$   & Full\\ \hline
$5$ & $\ge 60$- M &   $3$ vs  $4$ \& $5$ & E & $200$ & $100$   & Initial $20$  s\\ \hline
$6$ & $\ge 60$- M &   $4$ vs   $5$ & I & $100$ & $100$   & Initial $20$  s\\ \hline
$7$ &  M \&  F &   $1$ vs   $2$ & I & $100$ & $50$   & Full\\ \hline
$8$ &  M \&  F &   $1$ vs   $2$ & U & $100$ & $50$   & Full\\ \hline
\end{tabular}
\end{table*}

\subsection{Hierarchical XGBoost with Glottal and Formant Features}

In contrast to end-to-end neural approaches, we explored a two-stage hierarchical (see Figure \ref{fig:block_diagram}) model using gradient boosted trees (XGBoost) enriched with expert-crafted speech features. This approach explicitly incorporates domain knowledge about dysarthric speech by extracting glottal pulse, extracted using the SEDREAMS algorithm \cite{drugman09b_interspeech}, and formant frequency features, alongside patient demographics (age group and gender), as inputs to the classifier.  %along with acoustic features—five formants and seven glottal pulse parameters .

\subsubsection{Feature Extraction} 
\begin{figure}[!htbp]
    \centering
    \includegraphics[width=0.5\textwidth]{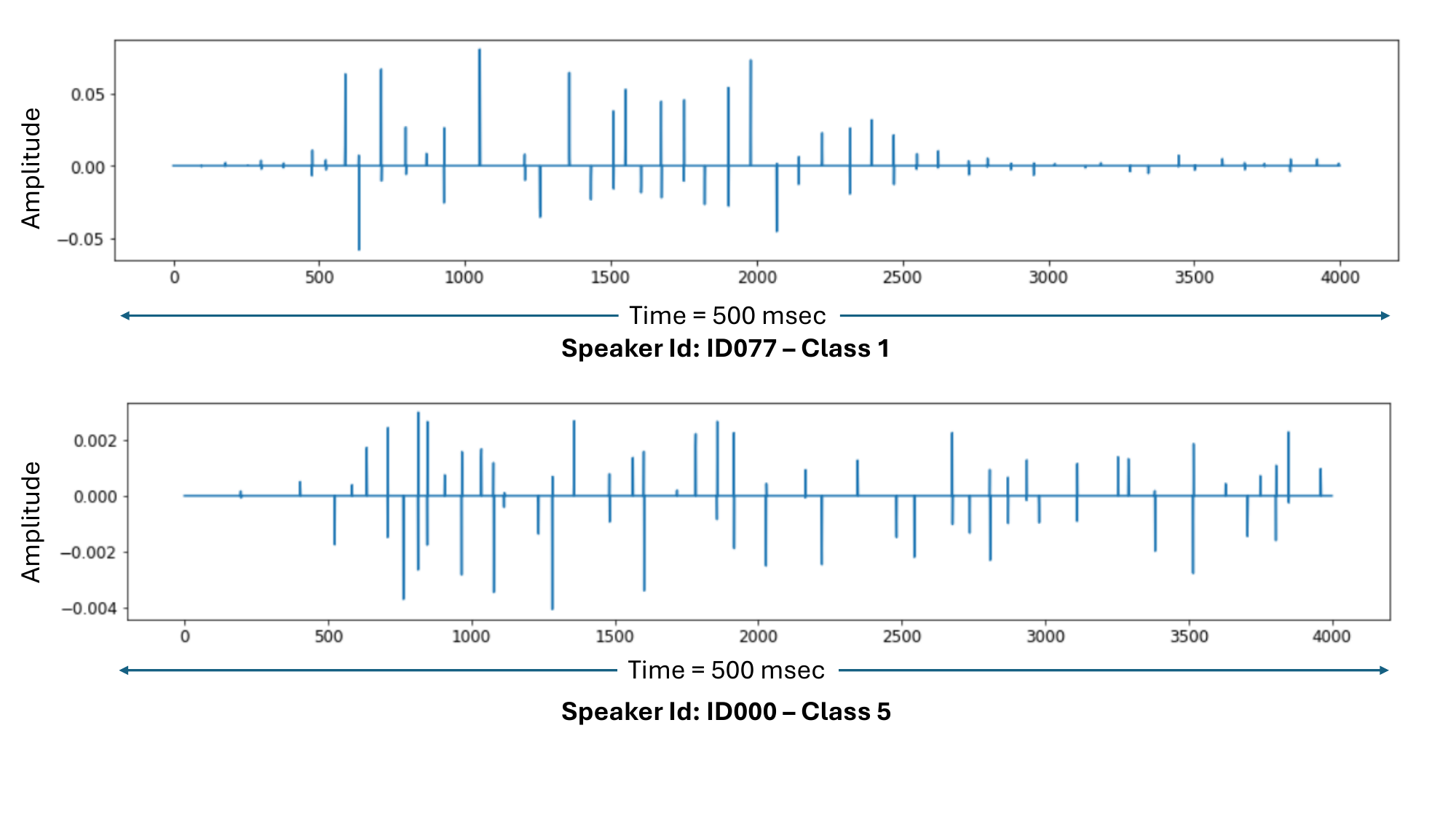}
    \caption{Glottal parameters extracted from speech signals for speakers $'ID077'$ (Class $1$) and $'ID000'$ (Class $5$).}
    \label{fig:glottal_patterns}
\end{figure}

From each audio recording, we computed 12 acoustic features that have known relevance in characterizing speech impairment: 5 vowel formant frequencies (F1–F5), and 7 glottal pulse parameters (see Fig. \ref{fig:glottal_patterns}). The glottal features were derived using the SEDREAMS algorithm to detect glottal closure instants (GCIs) – essentially markers of vocal fold vibration – from which measures such as pitch period statistics and amplitude are obtained. These features capture voice quality and articulation characteristics that may correlate with dysarthria severity (e.g., unstable pitch or reduced articulation clarity). In addition, two demographic features (speaker’s age group and gender) were included, since dysarthria manifestations can vary with age and differ between male and female speakers. 

\subsubsection{Architecture} 
\begin{figure}[!htbp]
    \centering
    \includegraphics[trim=0cm 13cm 13cm 0cm, width=1\linewidth]{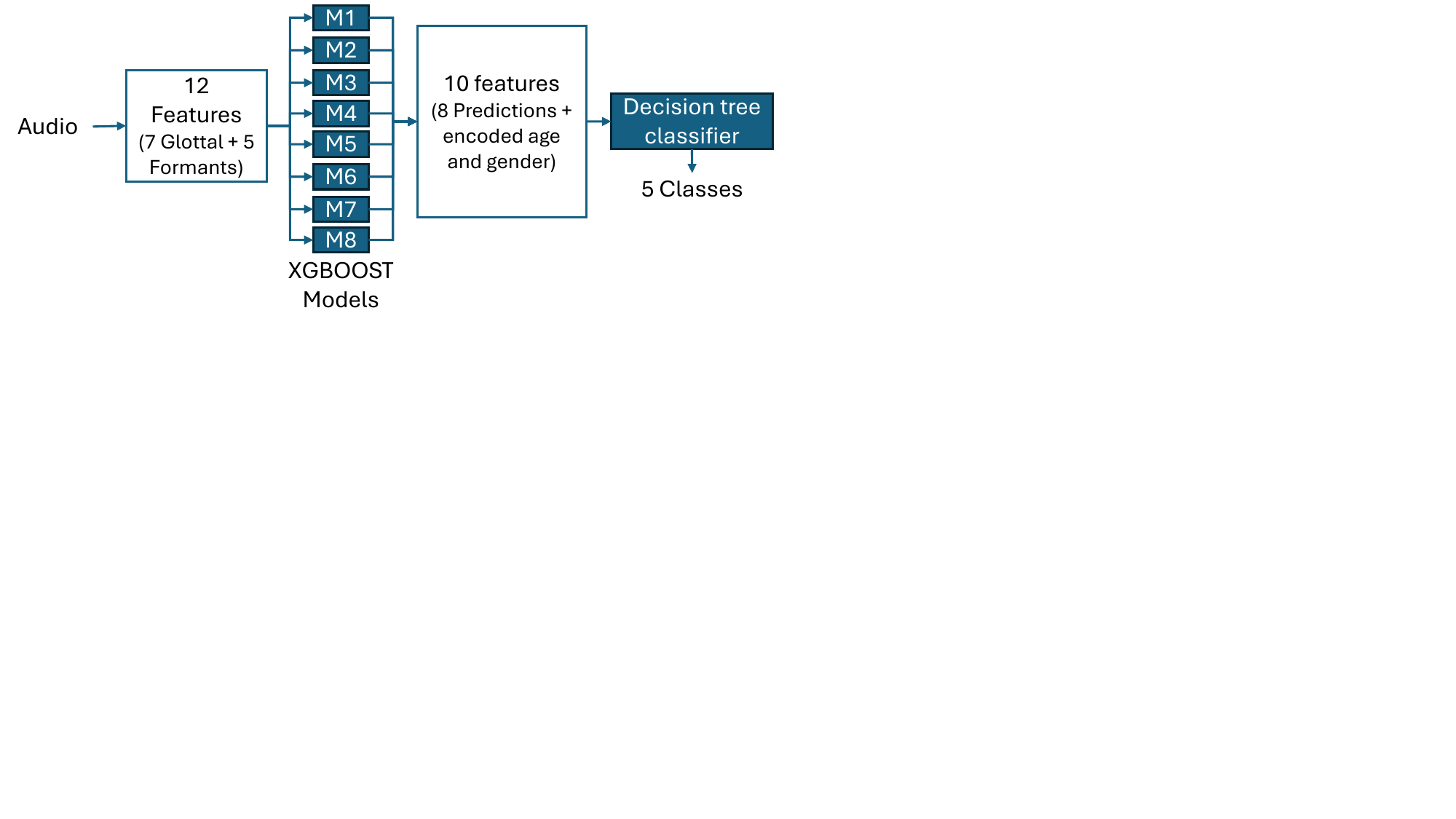}
    \caption{Block diagram of hierarchical XGBoost approach.}
    \label{fig:block_diagram}
\end{figure}

The classification is performed in two stages. In Stage 1, we train eight binary XGBoost classifiers (see Table \ref{tab:model_config}), each specializing in a particular decision sub-problem or subgroup of data. These were designed based on observed confusions and the metadata: for example, Model 1 distinguishes Class 3 vs Class 4 \& 5 for female speakers, Model 2 separates Class 4 vs Class 5 for female speakers, Models 3–6 handle similar binary splits for male speakers stratified by age (mid-age ($< 60$ vs old-age $\ge 60$), and Models 7 and 8 focus on differentiating lower severity classes (Class 1 vs Class 2) across all speakers. Each XGBoost model uses the 12 features as input, and all are trained with a small learning rate (0.01) to ensure fine-grained learning. The binary outputs from these 8 models (essentially predictions or confidence scores for each sub-task) are then collected. In Stage 2, we feed the collection of eight Stage-1 predictions plus the encoded age (normalized to [$0,1$]) and gender (encoded as $0.9$ for female and $0.1$ for male) into a higher-level classifier. Stage 2 is implemented as a simple decision tree (depth=5, 100 trees) that learns to map the pattern of binary outcomes to the final five-class decision. Figure \ref{fig:block_diagram} illustrates this hierarchy: the first layer of models focuses on specific binary distinctions, and the second layer integrates them for the overall classification. 

This hierarchical ensemble effectively performs a form of late fusion and decision fusion: it uses specialized classifiers to capture subtle differences (for example, differences in how females vs males present Class 4 vs 5), then combines their outputs. By incorporating demographics and dividing the task, it tackles the heterogeneous nature of the data (gender and age effects on speech) in a structured way. 

\subsubsection{Result} The hierarchical XGBoost approach yielded the highest performance among our methods. On the validation set of 53 speakers, it achieved a macro F1 score of 0.8644 (86.44\%), with an overall accuracy of ~0.88741. The confusion matrix indicated only slight confusion between adjacent classes (e.g., Class 3 vs 5, Class 4 vs 5). This high accuracy demonstrates the effectiveness of combining expert features with a tailored classification strategy. The model’s success suggests that in this limited-data scenario, incorporating prior knowledge (via features like formants and glottal pulses) can significantly boost performance. It’s worth noting that certain utterance types were excluded from this approach (the unvoiced PA and TA were found less useful for glottal feature extraction), indicating the approach made informed decisions about which data aspects to leverage. Overall, this method provides a strong benchmark, showing that a carefully designed feature-based model can excel in dysarthria classification.

% \textbf{Stage $1$:} Eight XGBoost binary classifiers with the parameters as shown in Table \ref{tab:model_config} are trained on subsets of data based on age and gender. Each model uses glottal parameters and formants for classification tasks such as distinguishing Class $3$ vs Classes $4$ \& $5$ or Class $4$ vs Class $5$.

% \textbf{Stage $2$:} Predictions from Stage $1$ are combined with encoded demographic features (age , gender  and fed into a decision tree ($100$ trees, max depth = $5$) for five-class classification.

% \textbf{Training:}  
% \begin{itemize}
%     \item XGBoost models: learning rate = 0.01, estimators = 100–200
%     \item Frame length: 50–500 ms depending on model
%     \item Decision tree: 100 trees, max depth = 5
% \end{itemize}

% \textbf{Feature Extraction:}  
% Glottal pulses are extracted using the SEDREAMS algorithm, which detects Glottal Closure Instants (GCIs) through a two-step process involving a mean-based signal and LP residual refinement. Seven glottal parameters are derived: number of peaks, and max/min/avg values for pitch duration and amplitude. Formants are obtained using the Parselmouth Python package.

\begin{my_note} 
% \textbf{Additional Details:}  
Separate models are used for male and female speakers because pitch and glottal characteristics vary significantly across gender and age groups. Experiments showed that phonation categories A, E, I, O, U and rhythm KA were most effective, while voiceless rhythms (PA, TA) were excluded due to poor glottal pulse extraction. 
% Models were trained using full audio, initial $20$ seconds, or last 10 seconds depending on optimal performance for each category.
\end{my_note}

\subsection{\onedcnnof{}: 1D-CNN with Output Fusion}

\begin{figure}[!htbp]
    \centering
    \includegraphics[width=0.45\textwidth]{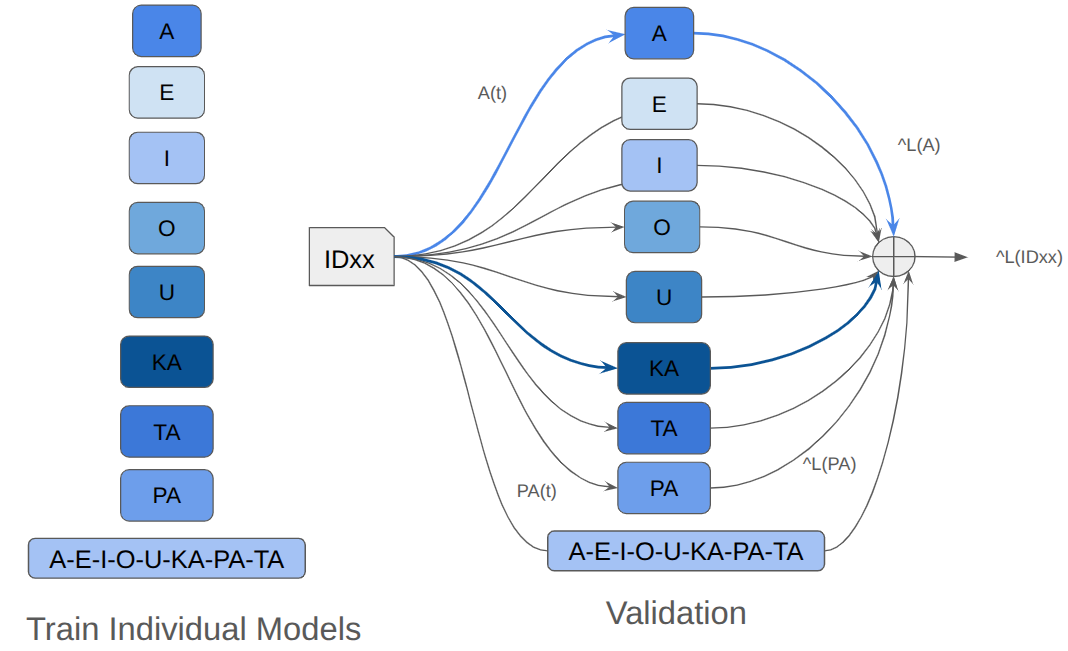}
    \caption{Architecture for late fusion model.}
    \label{fig:bilstm-of}
\end{figure}

The \onedcnnof{} approach (1-Dimensional Convolutional Neural Network with Output Fusion) addresses the classification by training separate CNN models for each type of utterance and then fusing their outputs via majority voting (see Fig. \ref{fig:bilstm-of}). The rationale is that each utterance (vowel or rhythm) might carry complementary information about the speaker’s speech deficit, so specialized models can learn features particular to each phonation, and a fusion decision then aggregates these insights. 

\subsubsection{Architecture \& Features} We built eight parallel 1-D CNN models, one for each of the 5 vowel phonations (A, E, I, O, U) and 3 rhythmic syllables (KA, PA, TA). Each CNN takes as input the raw audio of its designated utterance type, transformed into a sequence of acoustic feature frames. Unlike \vitof{} which used spectrogram images, \onedcnnof{} operates on frame-based features.

% Inputs are truncated/padded to $T=500$ frames ($\approx$ $5.01$ s) and processed by a 1D CNN stack:

We extracted a rich set of phase-based acoustic features per frame: specifically, 54-dimensional vectors comprising Phase Cepstral Coefficients  ($13$), Group Delay Cepstral Coefficients  ($13$), Modified Group Delay  ($13$), Instantaneous Frequency  ($13$), and other phase statistics like coherence and entropy. These features collectively capture fine details of the speech signal’s phase and frequency content (e.g., the shape of the spectrum, source-filter characteristics, and frequency modulation cues), which are informative for characterizing slurred or irregular speech in dysarthria. Each audio utterance was windowed (20 ms frames with 10 ms hop, 8 kHz sampling) and truncated or zero-padded to a fixed length of T = 500 frames ($\approx 5.01$ s) so that all inputs had equal length. 
The CNN architecture for each task is relatively small, namely,
\begin{enumerate}
    \item $3$ Conv1d layers with BatchNorm, ReLU, Dropout
    \item Global Average Pooling (across T frames)
    \item Layer Normalization
    \item Fully Connected Layers: $256$ $\rightarrow$ $128$ $\rightarrow$ $5$
\end{enumerate}
% roughly 3 convolutional layers (1-D conv with batch normalization and ReLU activation, followed by dropout for regularization), then a global average pooling across time frames, and 
The  fully connected layer maps to the 5 output classes. This yields a probability distribution over the 5 severity levels for that single utterance. Each of the 8 models has about 0.31 million parameters. We trained each CNN with a cross-entropy loss, using class-weighting inversely proportional to class frequency to handle the imbalance (so errors on rare classes counted more) and used the AdamW optimizer (learning rate 0.001). 
% No heavy pre-processing was applied to audio beyond trimming silence; the models learned directly from the raw waveform-derived features. To further alleviate class imbalance, data augmentation methods (such as perturbing audio or oversampling minority class examples) were employed during training, as mentioned in the paper. 

\subsubsection{Fusion Strategy} At inference time, a given speaker provides 8 utterances (A, E, I, O, U, KA, PA, TA). Each utterance is fed to its corresponding CNN model, producing 8 independent predicted labels. The final classification for the speaker is then obtained by majority voting across these 8 predicted labels. In other words, the class that appears most frequently among the eight CNN outputs is chosen as the overall diagnosis for that speaker. This late fusion by majority vote leverages the \textit{wisdom of the ensemble}.
%: if most utterance-specific models agree on a severity level, it is likely correct. It also 
This fusion starategy reduces the impact of any single utterance or model mis-prediction, as long as the majority are correct. This method assumes that each utterance contributes equally to the final decision and treats all output votes uniformly.

\subsubsection{Result} In initial experiments using all 8 utterance models, \onedcnnof{} achieved a validation accuracy of 0.6226 and a macro F1 of 0.5656. Analysis of feature importance using SHAP \cite{lundberg2017unified} indicated that certain feature types and certain utterances were more informative. By selecting the top 20 most important features and focusing on a subset of the best-performing utterance models (specifically using only the models for E, O, KA, PA, TA) while dropping weaker ones60, the performance improved significantly. The refined \onedcnnof{} system recorded a macro F1 score of 0.6398 (63.98\%) on the validation set, with a weighted F1 of 0.63. This improvement highlights that not all utterances contributed equally – the chosen five utterances had provided the most discriminatory power – and that feature selection helped remove noisy or redundant inputs. Even though \onedcnnof{}'s accuracy ($\approx$ 64\%) was lower than the transformer and XGBoost models, it demonstrated a viable approach using a straightforward architecture. The ensemble of simple CNNs was able to capture complementary information from different sounds, and the majority vote fusion yielded a stable combined decision. The approach underlined the importance of task-specific modeling (separating vowels and syllables) and feature engineering (phase features) for this problem, bridging a gap between end-to-end learning and expert features: the models learned internal representations, but on carefully chosen input features for each phonation. 

% Figure \ref{fig:bilstm-of} illustrates the late fusion architecture adapted for CNN-based modeling. This two-stage approach uses an ensemble of $5$ phonation and $3$ rhythm models, followed by majority voting fusion. Each CNN model is trained independently on utterances corresponding to its designated phonation or rhythm type, allowing it to learn discriminative features specific to that context. During validation, eight utterances per subject are processed by their respective models, and predictions are aggregated using majority voting to determine the final class label.

% \textbf{Preprocessing:}  
% No pre-processing was applied; raw audio recordings were directly utilized for feature extraction. We computed STFT with sampling rate $8$ kHz, FFT size $1024$, Hann window of $20$ ms, and hop of $10$ ms.

% \textbf{Feature Representation:}  

% \textbf{Feature Selection:}  
% SHAP-based analysis is applied to identify the top $20$ most influential features, improving model interpretability and performance.

% The model has $\sim$ $0.31$M parameters and was trained per task ($8$ tasks) using weighted cross-entropy (inverse class frequency), AdamW optimizer ($lr=0.001$), $100$ epochs, and early stopping (patience=$15$) based on macro-F1.

\subsection{\bilstmof: BiLSTM Ensemble with Output Fusion}

The \bilstmof{} approach extends the idea of utterance-wise modeling to sequence models, using Bidirectional LSTM (BiLSTM) networks for each utterance type and fusing their outputs with majority voting (see Fig. \ref{fig:bilstm-of}). Recurrent neural networks like LSTMs can capture temporal dynamics in speech, which might be beneficial for the rhythmic syllable utterances or any temporal patterns in phonation not captured by static features. 

\subsubsection{Architecture \& Preprocessing} We prepared the input for BiLSTMs similarly to \onedcnnof{}, but focusing on time-domain spectrogram features. All utterances were first trimmed to remove leading and trailing silences to ensure the model focuses only on the active speech portions. We then converted each utterance into a spectrogram using Short-Time Fourier Transform (STFT) with 20 ms window, 10 ms hop, and 256-point FFT (see Fig. \ref{fig:spectrogram-300ka}). The resulting magnitude spectrogram can be viewed as a sequence of frames (time steps) with 129 frequency-bin features each (given the 256 FFT, we have 129 unique frequencies). Unlike \vitof{}’s 2D image approach, here the spectrogram is fed as a time-series matrix to the LSTM. We constructed nine BiLSTM models: one for each of the same 8 utterances (A, E, I, O, U, KA, PA, TA) as in \onedcnnof{}, plus one extra model for a combined \textit{phonation-rhythm} ($\all$; concatenation of A-E-I-O-U-KA-TA-PA) 
% are independently trained using \textit{only} the samples corresponding to its designated phonation. This allows each model to learn discriminating features unique to that phonation context.
% 
input. 
\begin{my_note}
This combined input, denoted $\all$, is created by concatenating all 8 utterances of a speaker in sequence, forming a single longer waveform representing that speaker’s full set of sounds. The $\all$  model is intended to capture cross-utterance cues and overall speech characteristics when all content is heard together, complementing the individual utterance models. 
\end{my_note}

\begin{figure}[!htbp]
    \centering
    \includegraphics[width=0.225\textwidth]{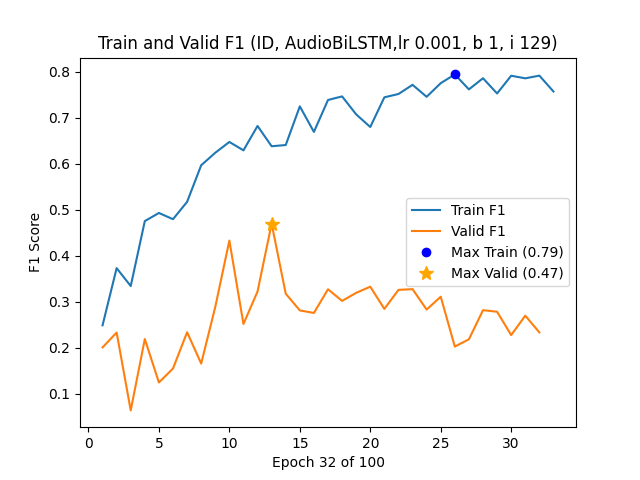}
    \includegraphics[width=0.225\textwidth]{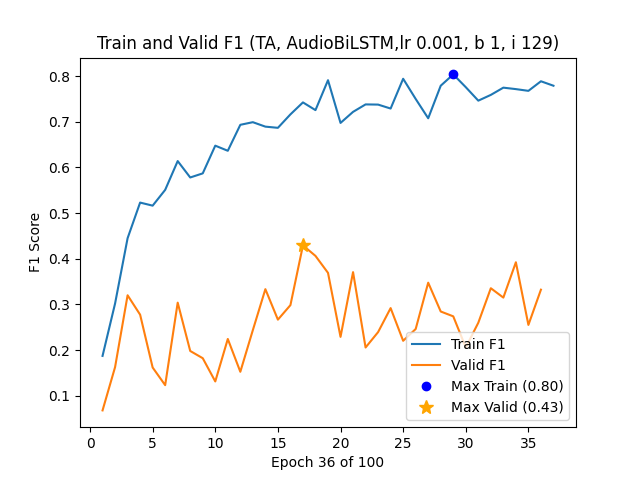}
    \caption{Sample train and validation F1-score versus epoch for $\all$ and rhythm TA.}
    \label{fig:all_train_valid}
\end{figure}

Each BiLSTM model is a lightweight recurrent network with an architecture as follows:
\begin{centering}
\em 
Input Sequence [N frames, $129$ features] (spectrogram)\\
$\downarrow$\\
Bidirectional LSTM Stack \\ ($4$ layers, hidden size $16$, dropout $0.1$) \\
$\downarrow$\\
Global Average Pooling (across N frames)\\
$\downarrow$\\
Layer Normalization \\
$\downarrow$\\
Dropout ($0.1$) + Fully Connected Layer (Linear)\\%, output: 5 classes) \\
$\downarrow$\\
Output: Class Probabilities [$5$].\\
\end{centering}
% 4 layers of bidirectional LSTM (each with 16 hidden units and dropout 0.1) followed by global average pooling over time, then layer normalization, a dropout of 0.1, and a final fully connected layer to the 5 classes. 
This yields about 31.6k trainable parameters per model. The design ensures the model can capture temporal patterns within each utterance (via the BiLSTM’s recurrent connections) while eventually producing an utterance-level class probability. We used cross-entropy loss with class weighting (inversely proportional to class frequency, as in \onedcnnof{}) to address the imbalance. Training used the Adam optimizer (lr=0.001) and early stopping (patience 20 epochs) to prevent overfitting. Because the dataset is small, we used a batch size of 1 %(each batch = one speaker’s one utterance) 
to maximize the training samples usage per epoch.  Figure \ref{fig:all_train_valid} shows a sample train and the validation plot 
%in terms of F1-score plots 
for \model{$\all$} and \model{TA}. 
As can be observed the F1-score is $0.47 (0.43)$ on the validation set achieved after $14 (17)$ epochs respectively for \model{$\all$} and \model{TA}.
% Data augmentation similar to the CNN case was applied to improve model generalization. 

\subsubsection{Fusion Strategy} During inference, a speaker’s 8 trimmed utterances are run through the corresponding 9 BiLSTM models (for A, E, I, O, U, KA, PA, TA, $\all$) to produce 9 predicted labels. 
% Additionally, we run the concatenated utterance ($\all$) through its BiLSTM, giving a 9th predicted label. 
The final classification is determined by majority vote among these 9 outputs like in case of  \onedcnnof{} (with one extra vote coming from the $\all$  model). 
% Essentially, each model “votes” on the speaker’s class, and the class with the most votes wins. This is similar to’s fusion but . The inclusion of $\all$  can be seen as a way to incorporate holistic information about the speaker’s voice beyond the isolated contexts of each sound. 

\subsubsection{Result} Initially, using all 9 BiLSTM models, the system achieved about 62.26\% macro F1 (very similar to the baseline CNN ensemble). We observed that certain models (particularly those for the sustained vowels) were less accurate, whereas the models for the rhythmic syllables (KA, PA, TA) and the combined utterance $\all$  were more reliable. Therefore, in a refined experiment, we pruned the ensemble down to 4 models – using only the three rhythm utterance models (KA, PA, TA) and the $\all$  model, which were the strongest contributors. With the majority voting among just these four outputs, the performance improved markedly. The best \bilstmof{} configuration achieved a macro F1 score of 0.7042 (70.42\%) on the validation set, with a weighted F1 of 0.6450 and accuracy of $\approx$ 0.642. This was a substantial gain over using all utterances indiscriminately, indicating that for LSTM models the rhythmic utterances carried more consistent signals of severity (possibly because these fast repetitive syllables exaggerate speech impairments like slurring or timing irregularities). The result of 70\% F1 also surpassed the \onedcnnof{}’s 64\%, suggesting that capturing temporal dynamics via BiLSTM (and including the combined utterance) provided an edge. However, \bilstmof{} did not reach the performance of the feature-based XGBoost model, underlining that purely data-driven approaches faced challenges with the limited data. Nonetheless, \bilstmof{} demonstrates an effective strategy: by tailoring models to different utterance types and wisely combining them, one can boost classification performance. 

\subsection{Results and Comparison} 

All four approaches were evaluated under the same conditions on the SAND validation set (53 speakers, true labels unknown to models during training). 

\begin{table*}[!htbp]
\centering
\caption{Confusion matrices for different approaches: (a) \vitof{}, (b) \onedcnnof{}, (c) \bilstmof{}, and (d) XGBoost.}
\begin{subtable}[t]{0.224\textwidth}
\centering
\caption{\vitof}
\begin{tabular}{cc|ccccc}
 & $\hat{L}$ & 1 & 2 & 3 & 4 & 5 \\
\hline
$L$ & 1 & 2 & - & - & - & - \\
    & 2 & - & 2 & 2 & - & - \\
    & 3 & - & - & 8 & 1 & 2 \\
    & 4 & - & - & 4 & 6 & 4 \\
    & 5 & - & - & 5 & 2 & 14 \\
\end{tabular}
\end{subtable}
\hfill
\begin{subtable}[t]{0.224\textwidth}
\centering
\caption{\onedcnnof{}}
\begin{tabular}{cc|ccccc}
 & $\hat{L}$ & 1 & 2 & 3 & 4 & 5 \\
\hline
$L$ & 1 & 1 & 1 & - & - & - \\
    & 2 & 1 & 3 & - & - & - \\
    & 3 & - & 1 & 7 & - & 4 \\
    & 4 & - & 3 & 1 & 5 & 5 \\
    & 5 & - & 2 & - & 2 & 17 \\
\end{tabular}
\end{subtable}
\hfill
\begin{subtable}[t]{0.224\textwidth}
\centering
\caption{\bilstmof{}}
\begin{tabular}{cc|ccccc}
 & $\hat{L}$ & 1 & 2 & 3 & 4 & 5 \\
\hline
$L$ & 1 & 2 & - & - & - & - \\
    & 2 & - & 2 & - & 2 & - \\
    & 3 & - & - & 7 & 3 & 2 \\
    & 4 & - & - & 1 & 9 & 4 \\
    & 5 & - & - & 3 & 4 & 14 \\
\end{tabular}
\end{subtable}
\hfill
\begin{subtable}[t]{0.224\textwidth}
\centering
\caption{XGBoost}
\begin{tabular}{cc|ccccc}
 & $\hat{L}$ & 1 & 2 & 3 & 4 & 5 \\
\hline
$L$ & 1 & 1 & - & - & 1 & - \\
    & 2 & - & 3 & - & - & 1 \\
    & 3 & - & - & 8 & 3 & 1 \\
    & 4 & - & - & - & 12 & 2 \\
    & 5 & - & - & - & 3 & 18 \\
\end{tabular}
\end{subtable}
\label{tab:combined_confmat}
\end{table*}

Table \ref{tab:combined_confmat} shows the confusion matrices for the four approaches while Table \ref{tab:sand_models} provides a comparison of the key characteristics and results of all the four approaches. We compare the model architecture, input features used, the fusion strategy to handle multiple utterances, and the achieved macro F1 score (the primary challenge metric) on the validation data. 

\begin{table*}[!htb]
\centering
\begin{tabular}{|c|p{3.0cm}|p{4cm}|p{4cm}|c|}
\hline
\textbf{Approach} & \textbf{Architecture} & \textbf{Input Features} & \textbf{Fusion Strategy} & \textbf{Macro-F1} \\
\hline
\vitof{} & ViT-B16 Transformer & Spectrogram images (512×256) & Average probability across 8 utterances & 0.68 \\
\hline
\onedcnnof{} & 1-D CNN (8-model ensemble) & Phase-based spectral features (54 dims/frame) & Majority vote across 8 model outputs & 0.64 \\
\hline
\bilstmof{} & BiLSTM (9-model ensemble) & STFT magnitude spectrogram (129 dims/frame) & Majority vote across 9 model outputs & \underline{0.70} \\
\hline
XGBoost & 8 XGBoost + Decision Tree & 5 formants + 7 glottal features + age/gender & Hierarchical ensemble (binary models $\rightarrow$ 5-class tree) & \textbf{0.86} \\
\hline
\end{tabular}
\caption{Comparison of four approaches for dysarthria severity classification in the SAND challenge.}
\label{tab:sand_models}
\end{table*}

% \subsubsection{Confusion Matrices}

\subsubsection{Observations}
 The feature-based Hierarchical XGBoost approach outperformed in terms of raw macro-F1 (0.86 vs 0.68–0.70 for the best neural models) which emphasizes the value of carefully chosen features and tailored sub-tasks in this challenge. By using formant and glottal features grounded in speech science, it appears to capture the relevant cues of dysarthria more directly, and the hierarchical structure effectively handled different speaker demographics and class confusions. On the other hand, the \vitof{} and \bilstmof{} approaches demonstrate that purely data-driven, end-to-end learning can also reach reasonable performance (0.68 and 0.70 F1 respectively) even with limited data, by exploiting augmentation, pre-training, or ensemble techniques. \vitof{}’s use of a pretrained transformer allows it to generalize from a small dataset by transferring knowledge from images, and its averaging strategy smooths out per-utterance variability. \bilstmof{} and \onedcnnof{} both emphasize the benefit of treating each utterance type separately.
%  – their confusion matrices showed that not all utterances are equally useful, and focusing on the most informative ones (like certain vowels or rhythmic syllables) improved their results119 120. 

It is interesting to note that both the fusion strategies employed, namely, averaging probabilities and majority voting have merit. \vitof{}’s probability averaging is a soft fusion, potentially leveraging the confidence of predictions, whereas \onedcnnof{} and \bilstmof{}’s majority vote is a hard decision fusion that can be more robust when individual model confidence is not well-calibrated. In our experiments, the majority vote (with selected strong models) in \bilstmof{} gave a higher F1 than \vitof{}’s averaging, but this may also be due to differences in feature learning capability (spectrogram image vs. sequential modeling). 

All models had to confront the class imbalance issue. The deep learning models did so through data augmentation and weighted losses, while the XGBoost model implicitly handled it by splitting tasks (ensuring minor classes got focused binary classifiers). As a result, each approach was able to identify Class 1 speakers (most scarce) to some extent.
% though performance on these fringe classes was understandably lower (confusion matrices from the papers show some confusion especially between class 4 and 5, and between class 3 and others123 124). 

% Despite the differences in absolute performance, we emphasize that these approaches are complementary rather than purely competitive. The ViT-AVE provides a quick transfer learning solution requiring minimal feature engineering, the CNN-OF and BiLSTM-OF provide interpretable per-utterance analysis (one can inspect which sounds were misclassified, gaining insight into the nature of the speech impairment), and the XGBoost model offers high accuracy with explainable feature importance (e.g., which formant or glottal feature was most decisive). In a real deployment, one might even consider an ensemble-of-ensembles – for example, combining the transformer and the XGBoost outputs – to further boost robustness, since they rely on very different signals (raw spectral patterns vs. hand-crafted features). 

% \section{Observations}
\begin{my_note}
% The data provided is highly imbalanced and it is biased towards classes $4$ and $5$. 
Automatic Speech Recognition techniques were unsuccessful on this dataset as the transcription (ASR output) provided are insufficient. For example, the rhythm sounds PA, TA and KA are repeated but there is no output transcription with information about how many times these sounds were repeated. If these rhythm sounds were annotated to indicate how many times each speaker uttered these, then ASR techniques could have been used to better advantage. Also, information about how the manual grading of the speech signals were done could have provided better background for designing systems. For example, we have assumed in all our experiments that the classification was arrived at by averaging over all the utterances. However exact methodology of manual classification would have provided a better background for the design of the loss function. 
\end{my_note}

\section{Conclusion}
We explored four complementary approaches for Task \#1 of the SAND Challenge, which involves five-class classification of dysarthria severity using speech recordings.
All the four approaches (\vitof{}, \onedcnnof{}, \bilstmof{}, and Hierarchical XGBoost) offer a unique perspective on the same classification task, from end-to-end neural modeling to hierarchical feature-driven classification. 
Comparative analysis shows that incorporating expert knowledge via features (formants, glottal pulses) can yield superior accuracy (macro F1 0.86), but deep learning methods with the right design (pre-training, data augmentation, ensemble fusion) can also achieve good performance (F1 0.70). The approaches are largely complementary: for instance, the transformer and LSTM models automatically learn abstract representations of dysarthric speech, while the XGBoost approach confirms the relevance of specific known biomarkers of dysarthria. All models face common challenges of limited and imbalanced data, which they address through augmentation, specialized modeling per utterance, and task decomposition. 

% In a unified paper, we compared and contrasted the 4 techniques so that 
Through this analysis, one can appreciate how different strategies effect the goal of classifying ALS speech severity. This collective insight suggests that future work could explore hybrid models that blend these approaches, such as feeding engineered features into neural networks or using neural outputs as features in boosting, to further improve robustness. Moreover, cross-validation with larger datasets or fusion of the complementary models could push performance beyond what each achieved. 
 
In summary, the SAND Task 1 challenge has been approached from multiple angles by us: from spectrogram-based deep learning to interpretable feature-based classifiers. The highest performing model (hierarchical XGBoost) demonstrates the effectiveness of domain-inspired features, while the deep learning models highlight the promise of end-to-end learning even in data-scarce settings. This consolidated study serves as a comprehensive reference for researchers interested in dysarthria classification, illustrating that there is no one-size-fits-all solution – instead, a combination of diverse techniques may be the key to robust and generalizable performance in this field.

\bibliographystyle{IEEEbib}
\bibliography{mybib}
\end{document}